\documentclass[10pt,journal,cspaper,compsoc]{IEEEtran}

\usepackage{amsmath, amssymb, enumerate}
\usepackage{fancybox}
\usepackage{amsfonts}
\usepackage{algorithm}
\usepackage{algorithmicx}
\usepackage{algpseudocode}
\usepackage[usenames]{color}
\usepackage{listings}
\usepackage{graphicx,dblfloatfix}
\usepackage{times}
\usepackage{psfrag}
\usepackage{subfigure}
\usepackage{caption}
\usepackage{multirow}
\usepackage{epsfig}
\usepackage{url}
\usepackage{color}
\def\fixme#1{\typeout{FIXED in page \thepage : {#1}}
\bgroup \color{red}{[FIXME: {#1}]} \egroup}
\usepackage{rotating,tabularx}

\begin{document}

\title{A Covert Channel Based on Web Read-time Modulation}
\author{
Joshua Davis\\
jmd@covert.codes\\
}

\maketitle
\thispagestyle{empty}

\begin{abstract}
A network covert channel is created that operates by modulating the time between web resource accesses, with an ``average web user'' read-time used as a reference.  While the covert channel may be classified as timing based, it does not operate by changing deterministic protocol attributes such as inter-packet delay, as do most timing based network covert channels.  Instead, our channel communicates by modulating transaction level read-time, which in the web browsing case has significant non-deterministic components.  The channel is thus immune to methods typically used to detect timing based network covert channels.
\end{abstract}

\section{Introduction}
Network covert channels are divided into storage based and timing based channels \cite{TCSEC}.  Storage based channels communicate their illicit data by inserting it into unused packet headers, or by modulating existing header data \cite{DeVivo98}.  Timing based channels operate by changing some aspect of the legitimate ({\em carrier}) communication timing, such as inter-packet delay \cite{Girling87}.  Timing based channels have historically relied on modification of lower level protocol timing aspects, with modulation occurring for example at the network or transport layers (see \cite{Luo12} for a recent example).  Timing aspects modulated by such channels may be quite deterministic, so detection has often focused on statistical analysis of the carrier communication \cite{Cabuk04} \cite{Cabuk09} \cite{Berk05}.  The channel derived in this paper is different than other timing channels in that it operates at a higher level protocol: the Hypertext Transfer Protocol (HTTP) as used in web browsing.  In a typical web browsing scenario, the timing between web page accesses (web {\em read-time}) is somewhat non-deterministic.  We will take advantage of this, modulating the read-time to send our covert message.

\section{Related Work}
The impetus for this work comes from recent work on a web based covert channel developed in \cite{Davis14} and \cite{Davis14_2}.  There, a covert channel was derived that communicated through web resource access ordering.  Web resource names were mapped to codes, and the codes of the covert message were transmitted by accessing the resources associated with the message codes, in the order the codes appeared in the message.  A fundamental aspect of that channel was user timing behavior emulation.  It is unrealistic to expect a user to access web resources with minimum or uniform delay between accesses.  The channels in \cite{Davis14} and \cite{Davis14_2} attempted to approximate a typical web browsing user's read-time by utilizing a formula provided by the Nielsen Company \cite{Nielsen08} that associates a web page word count with average read-time.  In that channel, the average yielded by the Nielsen formula was used as the mean for an exponential distribution, and the read-time was taken pseudo-randomly from that distribution.  The investigation into web read-time in \cite{Davis14} and \cite{Davis14_2} provided the impetus for the channel here derived, which operates entirely by modulating web read-time.

\section{User Web Transaction Behavior}
Accurate user behavior emulation involves more than simply choosing pseudo-random values from an exponential distribution.  In light of more advanced research in the field \cite{Liu10} \cite{Huang10} \cite{Weinreich06}, and observation (the generated read-times in \cite{Davis14} and \cite{Davis14_2} were much longer than would be expected for a typical user), it is apparent that the method used in those papers is overly simplistic.  For example, \cite{Liu10} shows that web read-times depend on many attributes of the page in question, including not only word count but also page geometry and keywords.  That paper also demonstrates that web read-times closely conform to Weibull distributions, of the sort used in reliability analysis.  For tractability, we will continue to use an exponential distribution (which, notably, is a Weibull distribution with a shape parameter of one) in the channel developed here, though we now utilize a scaling parameter for tuning.  The reader is urged to reference \cite{Liu10}, \cite{Huang10}, and \cite{Weinreich06} when developing a behaviorally accurate implementation of the channel discussed here.

However one models average web transaction behavior, the dynamic nature of the web makes prediction of read-time difficult in uncontrolled environments.  This is due in part to the dynamics of user behavior, which may be largely related to the web content.  Consider for example a page consisting of a news article.  The content may be short and condensed, with a few keywords that may aid in assessing the pages relevance to a given user or set of users.  However, many news sites include a comments section below the main page content, which may be very lengthy with respect to the article itself, and which may incidentally contain any number of keywords that may confuse a programmatic content assessment.  Also, we have found, in the course of developing our channel, that page size may fluctuate between accesses.  This is due to dynamic site content such as advertisements (which may also change geometry and keyword content of a page).  Further, read-time prediction is prohibitively difficult when the content is not primarily text based.  Pages containing videos and interactive content like games and chat rooms possess their own read-time dynamics.  For tractability, our test implementation for the channel derived here utilized web resources that were primarily text based, and that included little or no dynamic content.

\section{A Covert Channel Based on Web Page Read-Time}
Our transmitter communicates its message by altering web read-time, with degree of deviation from a pseudo-randomized and shared (between the transmitter and receiver) baseline communicating the code or codes to be sent between each set of web page accesses.  System delay, the largest portion of which is likely to be network delay, is significant in this system.  Since such delay is variable, the transmitter and receiver share knowledge of a {\em maximum system delay} parameter $\delta$.  Maximum system delay will probably not be known exactly, but a safe ceiling can be agreed upon.  $\delta$ should exceed the time required for the transmitter to perform a DNS look up, contact the remote web server, and retrieve the web resource, in the worst expected case.  As we will soon see, there are constraints that keep us from defining $\delta$ to be too large in our implementation, and there are benefits to using lower values.

Along with the maximum system delay, the transmitter and receiver share a minimum code-space delimiter $\alpha$, a scale factor $s$ to be used in conjunction with the formulated exponential distributions, and a Pseudo-Random Number Generator (PRNG) seed.  With this information shared between the communication endpoints, the transmitter is able to convey its covert message to the receiver via its web read-time behavior.

\subsection{Channel Transmitter}
The transmitter software operates thus.  A list of Uniform Resource Locators (URLs) is formulated or retrieved.  In \cite{Davis14} and \cite{Davis14_2}, this was done by having a daemon reside on the transmitting user's computer that continually monitored network traffic, extracting URLs from the pages as the user casually browsed the web.  The transmitter seeds its PRNG with the shared seed, ensuring that the transmitter and receiver maintain pseudo-random variable synchronization.  The transmitter selects pseudo-random URLs from its list until its entire message has been conveyed.  After each web page is retrieved, the average read-time $\mu$ is calculated using the Nielsen formula \cite{Nielsen08}, shown in Equation 1, where $w$ is the number of {\em relevant} words contained in the web page.  Word relevance is implementation specific; in our test implementations, words that were not HTML or JavaScript were considered relevant.  Accurately determining relevance may be difficult.  For example, many discussion posts at the bottom of a web page will significantly alter, presumably erroneously, the time an average user may be expected to spend on the page, should the transmitter consider the words in these posts relevant.  The average wait time $\mu$ may be scaled as shown in Equation 2, where $s$ is the scale factor, which is shared between the transmitter and receiver.

\begin{equation}
\mu = 0.44 \cdot w+25
\end{equation}

\begin{equation}
\mu_s = s\mu
\end{equation}

In the case of page access errors such as the familiar {\em 404 Not Found} error, the transmitter waits a small amount of time, pseudo-randomly selected between $0$ and $\alpha$, before accessing the next URL.  Equation 3 shows how this wait time is calculated in our test implementation, $r$ being a pseudo-random number evenly distributed between zero and one, taken from the seeded PRNG, and $t_r$ in seconds.  The use of $100$ to scale $r$ is arbitrary.

\begin{equation}
t_r = (100r \bmod \alpha) + \delta
\end{equation}

If no error condition exists upon accessing the web resource, a pseudo-random read-time is selected from a distribution formulated with $\mu_s$ as its mean, as shown in Equation 4, with $\rho$ in seconds and $r$ provided by the PRNG.  The standard deviation $\sigma$ of an exponential distribution is equal to its mean.  We limit the maximum transmission time to $\beta$, introduced in Equation 5, and measured in seconds.  By this definition, we limit our read-times to two standard deviations, shifted up by $\alpha$ (read-times between $0$ and $\alpha$ indicate an error).  Read-times may occur after $\beta$ seconds, but these are reserved for {\em bad codes}, which will be discussed below.  The choice of two standard deviations (plus $\alpha$) for $\beta$ in our test implementation is largely arbitrary.

\begin{equation}
\rho = -\mu_s \cdot \ln(1 - r)
\end{equation}

\begin{figure*}[ht]
  \centering
    \includegraphics[width=0.70\textwidth]{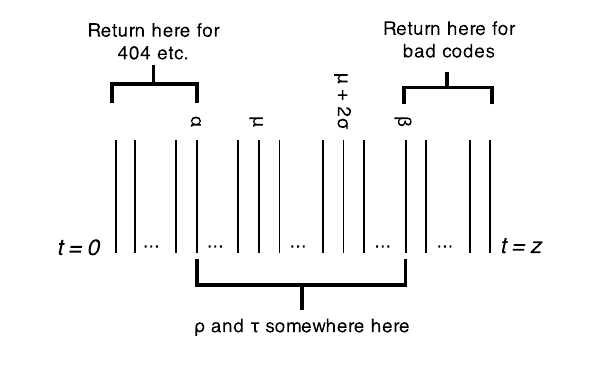}
  \caption{Visualization of System Time Domain.  Each interval (delineated by vertical lines) is $\delta$ seconds.}
\end{figure*}

% equation 5
\begin{equation}
\beta = \mu_s + 2 \sigma + \alpha = 3 \mu_s + \alpha
\end{equation}

$\nu$ is defined in Equation 6 as the total time-space around two standard deviations of the mean, divided by the number of codes in the code-space.  Our available time window for code transmission is divided evenly among the codes, and $\nu$ represents the size of the resulting intervals.  A number representing a code offset from $\rho$ is calculated as shown in Equation 7, with $\Omega$, and $\nu$ in seconds.  The variable $c$ represents the code offset, $0 < c < \lambda$.  Each $nu$ seconds after $\rho$ delineates a time interval representing a new code.  There are $\lambda$ codes in the code-space, so $0 <= \Omega <= \nu \cdot (\lambda - 1)$.

%equation 6
\begin{equation}
\nu = \frac{\beta - \alpha}{\lambda} = \frac{3\mu_s}{\lambda}
\end{equation}

%equation 7
\begin{equation}
\Omega = c \cdot \nu
\end{equation}

The final read-time is calculated as shown in Equation 8, with $\tau$ in seconds.  Note that a modulo occurs that wraps sufficiently high codes around to the beginning of the time-space, so the codes are split above and below $\rho$.  The transmitter will wait $\tau$ seconds before accessing another URL.  The final read-time $\tau$ allows the receiver to decode the the sent message character.  The transmitter will repeat this process until its entire message has been sent.  For messages $i$ codes long, the transmitter will access $i$ web resources in order to convey its message.

%equation 8
\begin{equation}
\tau = ( (\rho - \alpha + \Omega) \bmod (3\mu_s) ) + \alpha
\end{equation}

A {\em bad code} exists when $\rho > \beta$ (the proposed read-time is outside of two standard deviations above the mean, shifted up by $\alpha$), or $\nu < \delta$ (the size of code time slots is less than the maximum system delay.)  In the case of a bad code, the transmitter waits some time above $\beta$.  This extended read-time allows the receiver to easily identify bad codes, and ignore them.  The time that the transmitter waits in such cases may be constrained by an upper bound, identified as $z$ in our transmitter pseudo-code, which is Algorithm 1.  The time domain of our channel is shown visually in Figure 1.

\begin{algorithm}[h!]
  \begin{algorithmic}[1]
    \Function{transmit(array $msg$)}{}
      \State $D \gets $ URL list     \Comment E.g. from file
      \State $i \gets 0$
      \State seed PRNG			\Comment Seed shared with rcvr
      \vspace{0.25cm}
      
      \While {$i \le$ length($msg$)}
       \State $r \gets$ random()
        \State $w \gets$ words($D[random]$)
        \vspace{0.25cm}
        
        \If {$w = 0$}  \Comment Indicates 404 or other error
          \State sleep($\delta < t < \alpha$) \Comment Sleep $t$ in reserved zone
          \State continue
        \EndIf
        \vspace{0.25cm}
        
        \State $\mu_s \gets s(0.44w + 25)$
        \State $\sigma \gets \mu_s$
        \State $\rho \gets -\mu_s \ln(1 - r)$
        \State $\beta \gets \mu_s + 2 \cdot \sigma + \alpha$
        
        \State $c \gets msg[i]$
        \State $\nu \gets ( \beta - \alpha ) / \lambda$
        \vspace{0.25cm}
        
        \If {$\nu < \delta$ or $\rho > \beta$}  \Comment Bad code
          \State sleep($\beta + ( (100r) \bmod z $)
          \State continue
        \EndIf
        \vspace{0.25cm}
        
        \State $\Omega \gets c \cdot \nu$
        \State $\tau \gets ( (\rho - \alpha + \Omega) \bmod (\beta - \alpha) ) + \alpha$
        
        \State $i \gets i + 1$
        \vspace{0.25cm}
        
        \If{$i \le$ length($msg$)}
          \State sleep($\tau$)
        \EndIf
        \vspace{0.25cm}
      \EndWhile
   \EndFunction
 \end{algorithmic}
 \caption{Covert Channel Transmitter}
 \label{alg:algorithm1}
\end{algorithm}

\subsection{Channel Receiver}
The receiver must recover the covert message from the transmitter's read-time behavior.  As in other network covert channels, the receiver is assumed to have access to the transmitter communications in which the channel data is contained.  How this is done is scenario dependent; see \cite{Davis14_2} and \cite{Girling87} for some examples.  The receiver PRNG has been seeded with the same seed as the transmitter's, and it is assumed to be aware of the transmitter's read-times (each being $\tau$ for a resource access), and the word count of the pages the transmitter has accessed.  With the PRNG seeded, the receiver can synchronize pseudo-random numbers with the transmitter, though it must remember that the transmitter might use its PRNG for purposes other than deriving $\rho$ and compensate for this.  In our implementation the transmitter uses its PRNG to select pseudo-random URLs from its URL list; the receiver has been constructed with this in mind.

Having $\tau$ and $w$ for each page the transmitter has accessed, $r$ in synchronization with the transmitter, and the pre-shared values $\alpha$ and $\delta$, the receiver can derive $\mu$, $\sigma$, $\rho$, $\beta$, and $\nu$ the same way as the transmitter.  Recovering the message codes is then possible.  The receiver pseudo-code is shown as Algorithm 2.

\begin{algorithm}[h!]
  \begin{algorithmic}[1]
    \Function{receive()}{}
      \State $D \gets $ URL list     \Comment E.g. from file, includes times
      \State $i \gets 0$
      \State seed PRNG			\Comment Seed shared with receiver
      \vspace{0.25cm}
      
      \While {$i < $ length($D$)}
        \State $r \gets $random()
        \State $w \gets$ words($D[i]$)
        \State $\mu_s \gets s(0.44w + 25)$
        \vspace{0.25cm}
        
        \If{$i < $ length($D$)$-1$}
          \State $\tau \gets $time($D[i+1 -$time($D[i]$)
        \EndIf
        
        \If {$w = 0$}  \Comment Indicates 404 or other error
          \State $i \gets i + 1$
          \State $r \gets $ random() \Comment Due to site selection
          \State continue
        \EndIf
        \vspace{0.25cm}
        
        \State $\sigma \gets \mu_s$
        \State $\rho \gets -\mu_s \cdot \ln(1 - r)$
        \State $\beta \gets \mu_s + 2 \cdot \sigma + \alpha$
        \State $\nu \gets ( \beta - \alpha ) / \lambda$
        
        \If {$\nu < \delta$ or $\rho > \beta$}  \Comment Bad code
          \State $i \gets i + 1$
          \State $r \gets $random()
          \State continue
        \EndIf
        \vspace{0.25cm}
        
        \If{abs($\tau - \rho$) $< \delta$}  \Comment Absolute value function
          \State $\Omega \gets 0$
        \EndIf
        \If{$\tau > \rho$}
          \State $\Omega \gets \tau - \rho$
        \EndIf
        \If{$\tau < \rho$}
          \State $\Omega \gets (\beta - \rho) + (\tau - \alpha)$
        \EndIf
        \vspace{0.25cm}
        
        \State $c \gets $floor($\Omega / \delta$)  \Comment Next message code
        
        \State $i \gets i + 1$
        \State $r \gets $random()
        
      \EndWhile
   \EndFunction
 \end{algorithmic}
 \caption{Covert Channel Receiver}
 \label{alg:algorithm2}
\end{algorithm}

\subsection{Experimental Implementation}
In a test environment, we were able to transmit messages of varying length using the parameters $\alpha = 30$ seconds, $s = 1/4$, $\alpha = 30$ seconds, $\delta = 7$ seconds, and $\lambda = 32$ bytes.  The maximum message length tested was 14 codes.  The transmitter and receiver PRNGs were re-seeded at the beginning of each test.  The transmit URL list was specially selected to avoid interactive and dynamic content.

The receiver was positioned on the same network as the transmitter in such a way that it could observe the transmitter's HTTP traffic, and capture it using tcpdump.  The tcpdump output was used by the receiver script to recover the covert messages.  The transmitter and receiver software were both written in Python.

\subsection{Error and Data-rates}
Without specific knowledge about both the pseudo-random numbers that will be provided by the PRNG and the word count of the web pages that will be retrieved in the course of transmitting the message, it is prohibitively difficult to model the data-rate of our channel.  We can however arrive at some constraints that relate to the channel's data and error-rates.  First, a web resource retrieval will generate a {\em bad code} if $\mu_s < \frac{\lambda \delta}{3}$, so the mean read-time for a page must be above a threshold in order for the channel to communicate a code.  If a large number of URLs accessed by the transmitter have too low of a low word count, the channel error-rate will be high and its data-rate will decrease.  A bad code also occurs when $\rho > \beta$, which happens with certain combinations of $\mu_s$ and $r$.  One should keep in mind that the code-space size $\lambda$ and the maximum system delay $\delta$ have a direct effect on the error-rate, and not to set these parameters too high.

Errors notwithstanding, channel data-rate is proportional to $\delta$, the scale factor $s$, and is dependent on the pseudo-random numbers and page word counts encountered by the transmitter, as well as the message codes.  $\delta$ and $s$ are to some degree controlled by the user.  While it is necessary to hold $\delta$ high enough so that system latency will not occasionally overtake it, resulting in bad codes and wasted read-time, it is inefficient to set it {\em too} high, as the probability of bad codes increases with $\delta$.  Likewise, the scale factor should operate to assist in user behavior emulation, though setting it too low will increase the frequency of bad codes and decrease the channel data-rate.

\section{Comments and Future Work}
A significant challenge to our covert channel is the accurate emulation of user behavior (in terms of URL access ordering), a weakness also mentioned in \cite{Davis14} and \cite{Davis14_2}.  Users do not typically resources randomly, with no connection between hyperlink structure and subject matter.  A secure implementation of our channel should approximate natural resource ordering behavior.  One way such a system may be implemented is discussed in \cite{Davis14_2}.  There, it is suggested that the transmitter query a search engine with terms, then follow the top level result pages, traversing the resulting link structure coherently and semi-naturally.  Challenges remain, however.  For example, our test implementation spent a significant amount of read-time on a Terms of Use page it found on the web, which is presumably not typical user browsing behavior.

We reiterate that our read-time generation model is not accurate, and those who are interested in generating more accurate read-times should see \cite{Liu10}, \cite{Huang10}, and \cite{Weinreich06}.

In \cite{Davis14} and \cite{Davis14_2}, two codes were sent per web page access, increasing the data-rate of the channel significantly.  Presumably, such an improvement could be made to this channel as well, though as we have seen, increasing the number of codes in the channel's code-space results in further division of the available time-space, which increases the probability of a bad code ($\nu < \delta$).

The code-space about $\rho$ should be randomized in practice.  If a message consists primarily of high or low codes, our implementation would spend most of its read-time above or below $\rho$, which may increase the detectability of the channel.

\section{Conclusion}
We have created a network covert channel that transmits its message by modulating web resource read-time.  The channel utilizes a shared PRNG seed along with knowledge of ``average'' user read-time for a web page to formulate a base read-time.  Deviance from this base is used to convey the message codes.  Our tests have shown that the channel is functional given certain combinations of parameters.  This work is preliminary, and there is potential for extensibility in terms of both data-rate and security.

\bibliographystyle{plain}
\bibliography{reference}

\begin{thebibliography}{10}

\bibitem{Berk05}
Vincent Berk, Annarita Giani, George Cybenko, and NH~Hanover.
\newblock Detection of covert channel encoding in network packet delays.
\newblock {\em Rapport technique TR536, de lUniversit{\'e} de Dartmouth.
  Novembre}, 2005.

\bibitem{Cabuk04}
Serdar Cabuk, Carla~E. Brodley, and Clay Shields.
\newblock Ip covert timing channels: Design and detection.
\newblock In {\em Proceedings of the 11th ACM Conference on Computer and
  Communications Security}, CCS '04, pages 178--187, New York, NY, USA, 2004.
  ACM.

\bibitem{Cabuk09}
Serdar Cabuk, Carla~E. Brodley, and Clay Shields.
\newblock Ip covert channel detection.
\newblock {\em ACM Trans. Inf. Syst. Secur.}, 12(4):22:1--22:29, April 2009.

\bibitem{Davis14}
Joshua Davis.
\newblock A covert communication system using named resources.
\newblock Master's thesis, University of Kansas, April 2014.

\bibitem{Davis14_2}
Joshua Davis and Victor~S. Frost.
\newblock A covert channel using named resources.
\newblock 2014.

\bibitem{DeVivo98}
Marco de~Vivo, Gabriela~O. de~Vivo, and Germinal Isern.
\newblock Internet security attacks at the basic levels.
\newblock {\em SIGOPS Oper. Syst. Rev.}, 32(2):4--15, April 1998.

\bibitem{Girling87}
C.G. Girling.
\newblock Covert channels in lan's.
\newblock {\em Software Engineering, IEEE Transactions on}, SE-13(2):292--296,
  1987.

\bibitem{Huang10}
Jeff Huang and Ryen~W. White.
\newblock Parallel browsing behavior on the web.
\newblock In {\em Proceedings of the 21st ACM Conference on Hypertext and
  Hypermedia}, HT '10, pages 13--18, New York, NY, USA, 2010. ACM.

\bibitem{Liu10}
Chao Liu, Ryen~W. White, and Susan Dumais.
\newblock Understanding web browsing behaviors through weibull analysis of
  dwell time.
\newblock In {\em Proceedings of the 33rd International ACM SIGIR Conference on
  Research and Development in Information Retrieval}, SIGIR '10, pages
  379--386, New York, NY, USA, 2010. ACM.

\bibitem{Luo12}
Xiapu Luo, E.W.W. Chan, Peng Zhou, and R.K.C. Chang.
\newblock Robust network covert communications based on tcp and enumerative
  combinatorics.
\newblock {\em Dependable and Secure Computing, IEEE Transactions on},
  9(6):890--902, 2012.

\bibitem{Nielsen08}
Jakob Nielsen.
\newblock How little do users read?, 2009.

\bibitem{TCSEC}
United States~Department of~Defense.
\newblock {\em Trusted Computer System Evaluation Criteria}.
\newblock December 1985.

\bibitem{Weinreich06}
Harald Weinreich, Hartmut Obendorf, Eelco Herder, and Matthias Mayer.
\newblock Off the beaten tracks: exploring three aspects of web navigation.
\newblock In {\em Proceedings of the 15th international conference on World
  Wide Web}, pages 133--142. ACM, 2006.

\end{thebibliography}

\end{document}